\documentclass[twocolumn,amsmath,amssymb,aps]{revtex4}

\usepackage{graphicx}
\usepackage{dcolumn}
\usepackage{bm}

\usepackage{graphicx}
\begin{document}

\title{Bose glass behavior in (Yb$_{1-x}$Lu$_x$)$_4$As$_3$ representing the randomly diluted quantum spin-1/2 chains 
}

\author{G. Kamieniarz}
\affiliation{Faculty of Physics,
A. Mickiewicz University, ul. Umultowska 85, 61-614 Pozna\'n, Poland}
\affiliation{Max Planck Institute for the Physics of Complex Systems, 01187 Dresden, Germany}

\author{R. Matysiak}
\affiliation{Institute of Engineering and Computer Education,
University of Zielona G\'{o}ra, ul. prof. Z. Szafrana 4,
65-516 Zielona G\'{o}ra, Poland}

\author{P. Gegenwart}
\affiliation{Max Planck Institute for Chemical Physics of Solids, 01187 Dresden, Germany}
\affiliation{Experimental Physics VI, Center for Electronic Correlations and Magnetism, University
of Augsburg, 86159 Augsburg, Germany}

\author{A. Ochiai}
\affiliation{Center for Low Temperature Science, Tohoku University, 
Sendai 980-8578, Japan}

\author{F. Steglich}
\affiliation{Max Planck Institute for Chemical Physics of Solids, 01187 Dresden, Germany}

\begin{abstract}

The site-diluted 
compound (Yb$_{1-x}$Lu$_x$)$_4$As$_3$ 
is a scarce realization of the 
linear Heisenberg antiferromagnet partitioned into finite-size segments
and is an ideal model compound for studying field-dependent
effects of quenched disorder
in the one-dimensional antiferromagnets.
It differentiates from the systems studied so far in two aspects - the type of randomness and the nature of the energy gap in the pure sample.
We have measured the specific heat of single-crystal  (Yb$_{1-x}$Lu$_x$)$_4$As$_3$ in magnetic fields up to 19.5 T. The contribution $C_{\perp}$ arising from the magnetic subsystem in an applied magnetic field perpendicular to the chains is determined. Compared to pure Yb$_4$As$_3$, for which $C_{\perp}$ indicates a gap opening, for diluted systems a non-exponential decay is found at low temperatures which is consistent with the thermodynamic scaling of the specific heat established for a Bose-glass phase.

\end{abstract}

\maketitle

An exact correspondence between a quantum antiferromagnet and a lattice 
Bose gas was recognized \cite{Matsubara} much before Bose-Einstein condensation (BEC) was predicted in a three-dimensional array of antiferromagnetically
coupled ladders or dimers \cite{GiamarchiPRB1999,Rice2002,WesselPRL2001} and was experimentally observed \cite{NikuniPRL2002,Guedel} in the magnetic compound 
TlCuCl$_3$. 
In spin-gap systems, apart from the superfluid and Mott insulating states, a Bose-glass state
was also predicted \cite{FisherPRB1989}.
Since this observation in TlCuCl$_3$, a search for the BEC transitions and
a Bose-glass (BG) phase in magnetic materials 
has hastened 
\cite{GiamarchiNP2008,YuNature2012}, including also the spin-gap S=1 Haldane chains \cite{ZheludevPRL2002,MaedaPRL2007,ZvyaginPRL2007} or quasi-one-dimensional systems consisting of weakly interacting chains or two-leg ladders \cite{ManakaPRL2008,ManakaPRB2009}.

In spin-gap one-dimensional antiferromagnets (AFM) the ground state is 
a spin singlet 
separated by a finite energy gap $\Delta$ from the first excited state which is 
a spin triplet. In the integer-spin chains the gap $\Delta$ exists for 
uniform systems \cite{Haldane}, whereas in the case of the half-integer-spins, the gap may originate from the bond alternation. Then the pairs of strongly coupled spins exhibit the dimer singlet ground states which contribute to the ground state of the chain. However, the energy structure with a non-magnetic singlet 
ground state
is unstable for magnetic fields exceeding the critical value B$_c$ corresponding to $\Delta$ or for a sufficient level of dilution 
\cite{ManakaPRB2009}.

The BG state is an unusual state of matter with no broken symmetry and no energy gap in the excitation spectrum. This feature of BG was found for interacting bosons 
in quenched disordered systems \cite{ZvyaginPRL2007,ManakaPRB2009,YuNature2012,HongPRB2010}. The gapless nature of the BG state is characterized in the presence of an external magnetic field by an exponential magnetization behavior \cite{TRoscilde}, a finite uniform magnetic susceptibility and 
a non-exponential decay of the low-temperature specific heat \cite{YuNature2012}. 
The thermodynamic 
signature which uniquely characterizes the main features of 
the BG and Mott glass phases is a stretched exponential behavior of the specific heat \cite{YuNature2012}
given by the expression
\begin{equation}
 C(T) =  A \cdot (k_BT/J)^{-5/4} \exp(-\gamma(k_BT/J)^{-1/2}),
\label{Cv}
\end{equation}
where the parameters $A$ and $\gamma$ depend both on a concentration 
of impurities and on magnetic field, and $J$ represents a leading coupling 
constant. The unconventional magnetization and specific heat behavior of the BG state is elucidated by a local-gap model \cite{YuNature2012,TRoscilde,Haas}, considering the lowest order finite-size scaling of the emerging energy gaps. 

In the scenarios described so far, the spin gap is formed by the special 
{\it couplings} present in the quasi-one-dimensional (1d) compounds. 
For the integer spin $S=1$
Ni$^{2+}$-based compound \cite{YuNature2012},
the strong easy plane anisotropy ($D/J \gg 1$) is needed, whereas for the half-integer-spin compound with two-leg ladder structures embedded, strong rung-oriented and weak leg-oriented interactions are prerequisite
\cite{ManakaPRB2009,ManakaPRL2008} to get a pseudo $S=1$ Haldane chain representation. The role of the applied magnetic field is to close the 
spin gap and to create a finite density of bosons forming a magnetic BEC both in the pure and the doped compound \cite{YuNature2012}. 

However, the spin gap 
can be also induced by the transverse magnetic field applied in 
antiferromagnetic spin-1/2 chains with a Dzyaloshinskii-Moriya (DM) interaction
\cite{DenderPRL97,KoeppenPRL1999,Feyerherm}.
A good example of such a system is Yb$_4$As$_3$~\cite{Schmidt}. 
Upon dilution of Yb$_4$As$_3$ by the lutetium ions Lu$^{+3}$, 
the magnetic Yb$^{3+}$ is substituted by the
chemically identical non-magnetic Lu$^{3+}$, and the charge ordering 
in the site-diluted 
(Yb$_{1-x}$Lu$_x$)$_4$As$_3$  is retained for $x\leq 0.06$~\cite{Aoki}.  
The non-magnetic impurities are randomly distributed, 
giving rise to a statistical 
partitioning of the spin chains into even and odd-numbered segments
with the gaped singlet-triplet and spin-1/2 doublet ground states, respectively
\cite{Haas,KFA}. 
They result in a strong reduction of 
{\it zero-field} specific heat~\cite{prb2013rm} in the diluted Yb$_4$As$_3$. Moreover, the temperature dependence of the specific heat
fulfills~\cite{prb2013rm} the scaling law predicted 
for segmented Heisenberg spin chains~\cite{Haas}  
which is identical to Eq.~(\ref{Cv}) characterizing the BG state.

In this paper we aim at showing that the title compound 
(Yb$_{1-x}$Lu$_x$)$_4$As$_3$ provides a new type of the 
BG system under quenched doping by the non-magnetic impurities 
and subject to the transverse field. 
We argue that we observe this BG behavior because: a) the non-exponential
decay of the low-temperature specific heat is demonstrated both in the pure sample ($x$~=0\%) and in the site-diluted samples ($x$~=1\%, $x$~=3\%) if the applied field is absent; b) in the presence of magnetic 
field $B$ with a perpendicular component, 
the exponential decay occurs in the pure system only, while it becomes non-exponential
in the diluted systems;
(c) in the latter case the stretched exponential behavior (\ref{Cv}) of the specific heat characterizing the BG phase is validated.

The diluted Yb$_4$As$_3$ is very suitable for observation 
of the BG phase in quantum magnets. 
The diamagnetic dopands Lu$^{3+}$
create a site-diluted chains with missing adjacent bonds. 
The resulting system is a simple collection of finite linear segments.
Its analysis is void of further approximations, in contrast to a more complex physics of the systems 
with randomized bonds realized so far by a bromine doping~\cite{YuNature2012}
which affects locally
the values of couplings and anisotropy parameters and proliferates their number. 
Consequently,
the stretched exponential scaling law can be checked unambiguously because interactions in (Yb$_{1-x}$Lu$_x$)$_4$As$_3$ still depend on 
{\it a single} coupling constant J, and all the model parameters are fixed.

We have performed field-dependent measurements 
on the single-crystal sample (Yb$_{1-x}$Lu$_x$)$_4$As$_3$
with $x=1\%$ or $x=3\%$.
Both the samples and equipment were the same as those in our previous study
\cite{prb2013rm}. 
The magnetic specific heat $ C_{\text{m}}$ was obtained
from the measured specific heat $C_{\text {exp}}$ by subtracting 
the lattice contribution $C_{\text {ph}}$ estimated earlier~\cite{prb2009rm},
and assuming that
the phonon part is unaffected by the field applied and the doping
(see Fig.~1 in Ref. \onlinecite{prb2013rm} and Figs.~5-8 in Supplemental 
Material, in short SM). The heat capacity has been measured with error below 10\%. The error is largest at the lowest T and highest field.

The magnetic field has always been applied along one of the cubic [111] directions for each studied single crystal. The charge ordering transition selects one space diagonal as the spin chain direction and results in a polydomain state at low temperatures. Then in 1/4 of the domains the field is parallel to the magnetic chains, whereas in the remaining 3/4 of the domains, the chains will be subject to an effective perpendicular field~\cite{prb2009rm,prb2013rm}, so that
\begin{equation}
C_{\text {m}}(T,B)=0.75 \cdot C_{\perp}(T,B_{\text {eff}})+0.25 \cdot C_{\parallel}(T,B) \, ,
\label{specific_heat}
\end{equation}
where $B_{\text {eff}}=B \cdot \sin(70^{\circ})$ and  $C_{\perp}= C_{\parallel}$ in the limit $B=0$.
Here $C_{\parallel}$ and $C_{\perp}$ denote the magnetic heat capacity contributions from domains with the chains parallel and perpendicular to the applied field, respectively.

Following the consensus that the chains are well isolated and the XXZ 
anisotropy between the adjacent Yb$^{3+}$ ions is canceled by the DM 
antisymmetric exchange \cite{Schmidt,Shiba, Shibata},
the physical system 
is described by the one-dimensional effective spin-1/2 model 
\begin{eqnarray}
{\mathcal{H}} & = &  J \sum_{i=1}^{L}
{\mathbf S}_{i} \cdot {\mathbf S}_{i+1} -
g_{\perp} \mu_{B} B^{x} \sum_{i=1}^{L} S_{i}^{x}+
\nonumber \\
& - & g_{\perp} \mu_{B} B^{y}_{s} \sum_{i=1}^{L} \left( -1 \right)^{i}
S_{i}^{y} - g_{\parallel} \mu_{B} B^{z} \sum_{i=1}^{L} S_{i}^{z},
\label{hamilt}
\end{eqnarray} 
representing the chains pointing along the $z$ axis, subject to the perpendicular
$B_{\perp}$ or parallel $B^z$ field applied in the $x$ 
or $z$ direction, respectively. The lengths $L$ of the chains 
are assumed to be infinite
for the pure Yb$_4$As$_3$ compound and finite for doped samples.
Due to the DM interaction \cite{Oshik,Shiba}, the 
perpendicular field implies both
the renormalized field $B^{x}=B_{\perp}\cos(\theta)$ in the $x$ direction and
the staggered field $B^{y}_{s}=B_{\perp}\sin(\theta)$ in the $y$ direction,
where $\theta$ is the corresponding phase factor. 

In the model (\ref{hamilt}) all the parameters are fixed, and their values
arise from the earlier magnetic studies \cite{prb2009rm,Iwasa,Shibata}. 
We assume here that $J/k_B = 28$~K, $g_{\perp} = 1.3$,  $g_{\parallel} = 2.9$, 
$\tan \theta = 0.19$.
The applied field $B$ is matched 
to $B_{\perp}=B \sin(70^{\circ})$ and $B^z=B$ in our model. Its
thermodynamic properties are analyzed by 
the quantum transfer matrix (QTM) technique \cite{prb2009rm,prb2013rm,CMS2003}, having checked the fast convergence of approximants and the linear finite-size scaling of the specific heat in the presence of magnetic field, similar to that illustrated in Fig.~2 in Ref.~\cite{prb2013rm}. 

\begin{figure}
\begin{center}
\includegraphics*[scale=0.20,clip]{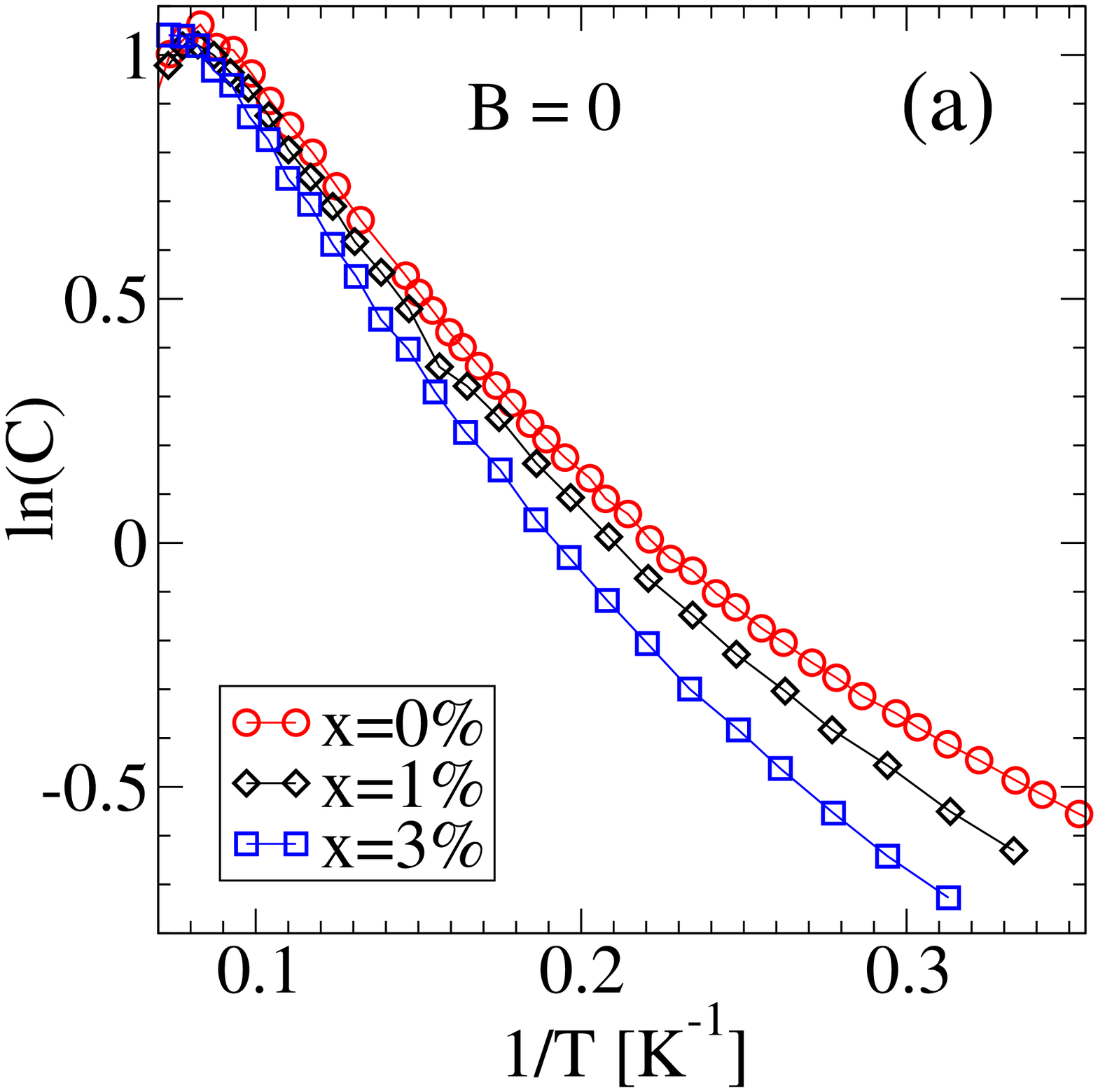}
\quad
\includegraphics*[scale=0.20,clip]{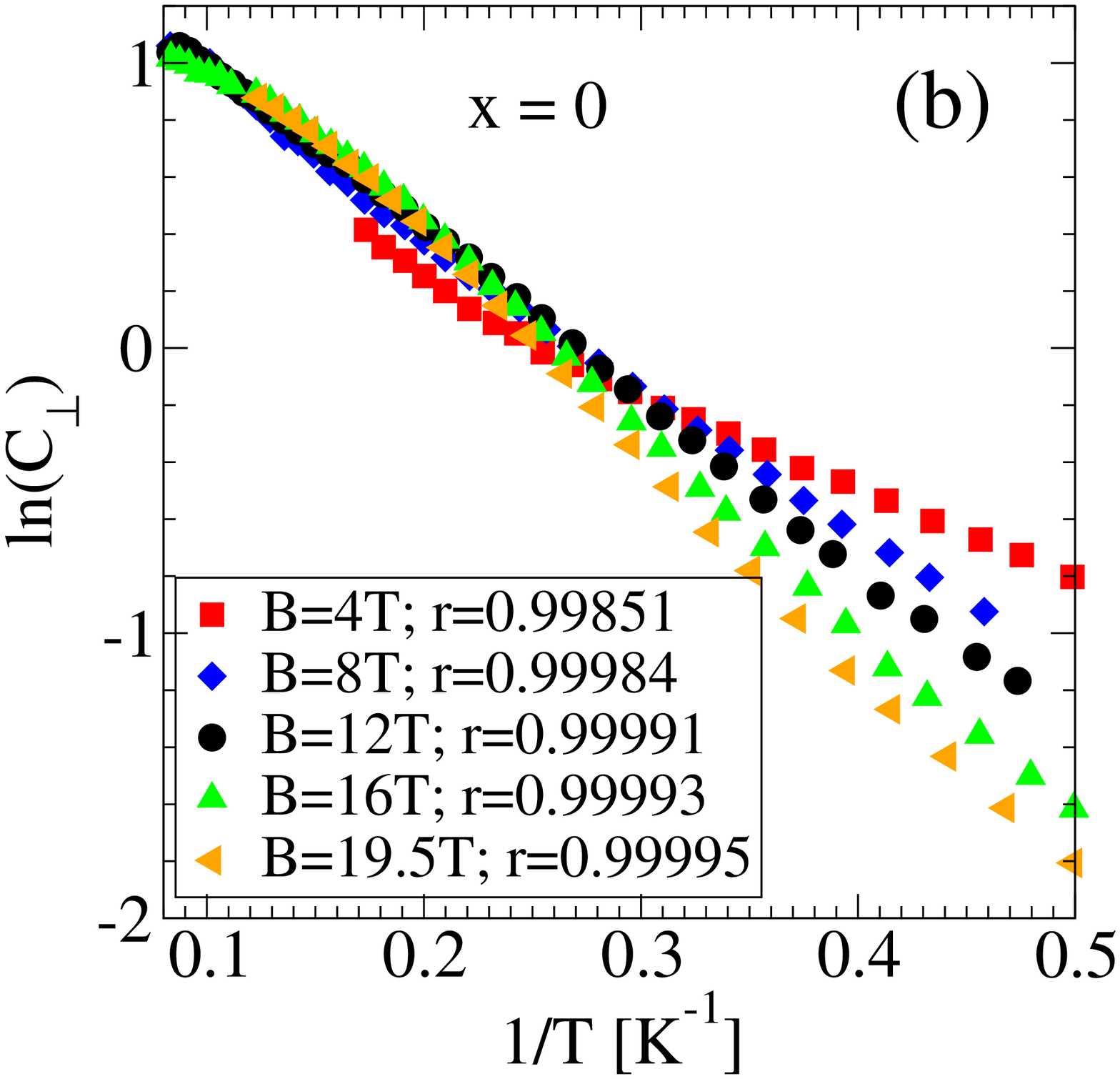}
\end{center}
\caption{(Color online) (a) Molar zero-field magnetic specific heat 
of the pure and doped samples 
plotted in semilogorithmic scale as a function of $1/T$. (b) The exponential dependence of $C_\perp$ 
as a function of $1/T$ in the low-temperature region. 
The corresponding Pearson correlation coefficients $r$ are provided.
}
\label{fig1}
\end{figure}

In Figs.~\ref{fig1}a and~\ref{fig1}b
some reanalyzed results for 
the pure~\cite{prb2009rm} Yb$_4$As$_3$ and the diluted~\cite{prb2013rm} (Yb$_{1-x}$Lu$_x$)$_4$As$_3$ are plotted. 
The {\it zero-field} data in Fig.~\ref{fig1}a demonstrate the non-exponential dependence of the molar specific heat 
on $1/T$ in the low temperature region regardless of the concentration $x$.
This feature is the manifestation of an essentially 
gapless energy spectrum. Even though in 
the diluted sample, the segments with even number of magnetic sites 
display residual energy 
gaps, their values are randomly distributed, their average values are small
for $x \le 0.03$ so that, instead the stretched exponential dependence is developed~\cite{prb2013rm,Haas}.

However,
in the pure sample subject to a field perpendicular to the chains ($B_{\perp} \ne 0, B^z=0$), the gap should open up and bring on an exponential
decay of the specific heat. 
The data $C_{\perp}(T)$ presented in Fig.~\ref{fig1}b are derived from 
the $C_{\text{m}}(T)$ results given by open symbols in Fig.~5 in Ref.
\onlinecite{prb2009rm} after subtracting according to Eq.~(\ref{specific_heat})
the contribution $C_{\parallel}$ 
from the domains with the chains aligned along the field which was
calculated, imposing $B_{\perp}=0$ in the model (\ref{hamilt}). The values $0.75 \cdot C_{\perp}$ extracted from $C_{\text{m}}(T)$ are very close to the corresponding QTM estimates plotted therein. In the relevant temperature region $2.5 \le T \le 5$~K the phonon part $C_{\text{ph}}(T)/T$ 
is much smaller than the contributions from $C_{\parallel}(T)/T$ and $C_{\perp}(T)/T$ shown in Fig.~5 in Ref.~\onlinecite{prb2009rm} so that phonons do not affect the accuracy of the extracted $C_{\perp}$.

The expected feature for the pure Yb$_4$As$_3$ is revealed in Fig.~\ref{fig1}b 
in the low temperature region, where the linear dependence on $1/T$ is
recovered for $C_{\perp}$.
The Pearson correlation coefficients {\it r}
measuring the linear correlation between two variables 
are sufficiently close ($0.99851 \le r \le 0.99995$) to the ideal value 1.
In addition, referring to Eq.~(\ref{Cv}), we have analyzed the non-exponential dependence of $C_{\perp}$ which leads to a deterioration of the linearity ($0.99814 \le r \le 0.99874$, see Fig.~5 in SM).

\begin{figure}
\begin{center}
\includegraphics*[scale=0.20,clip]{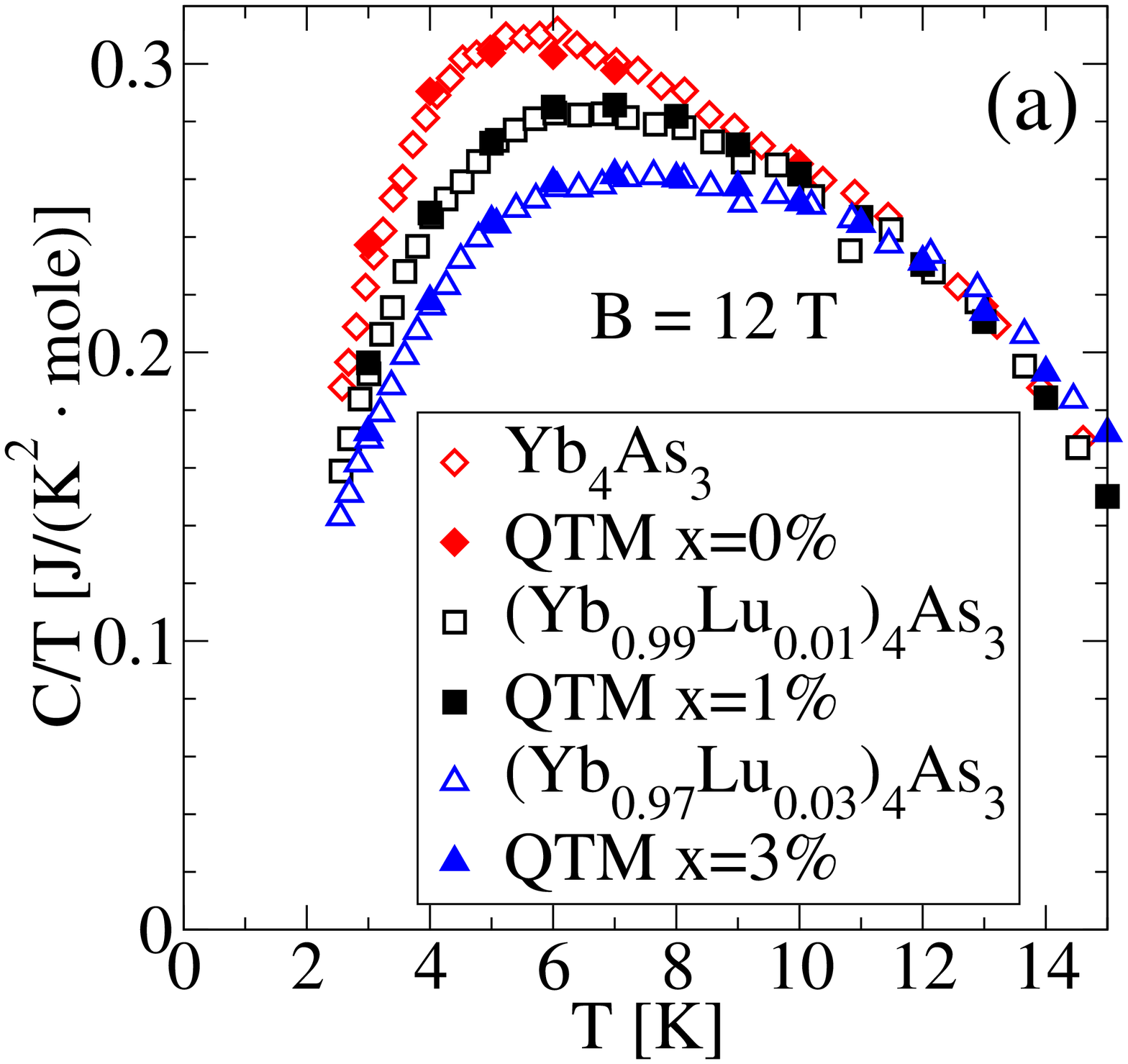}
\includegraphics*[scale=0.20,clip]{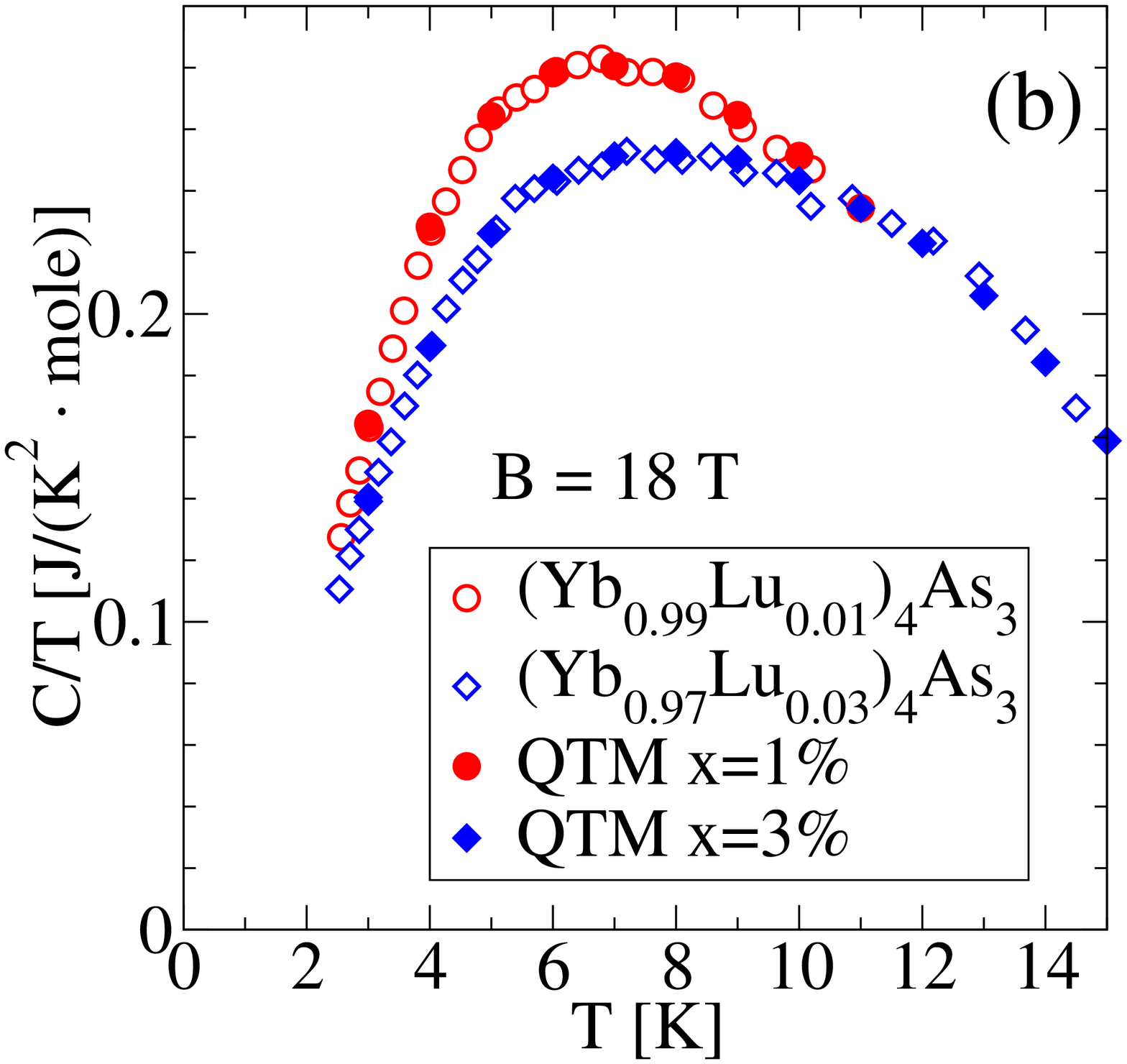}
\includegraphics*[scale=0.20,clip]{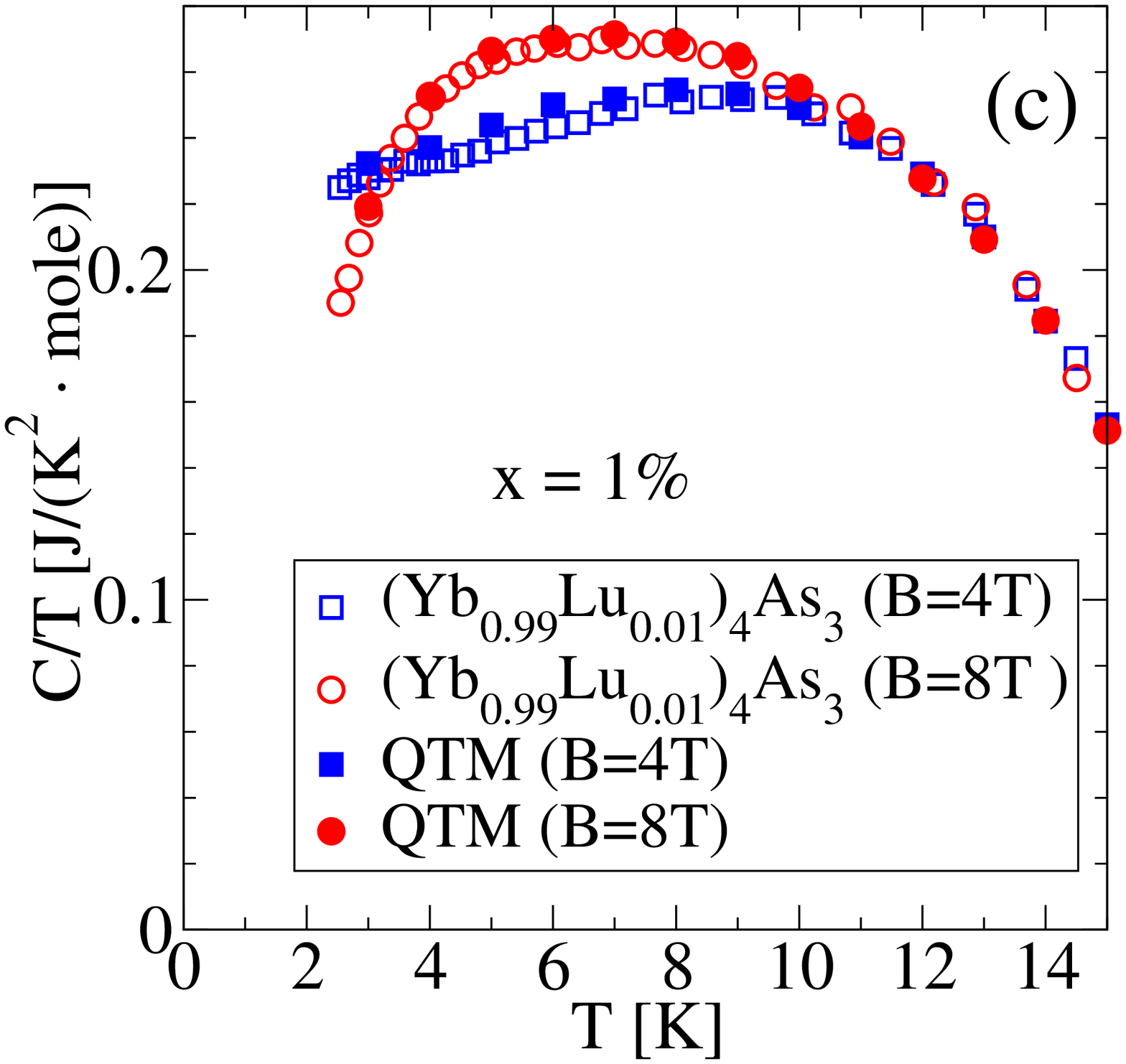}
\includegraphics*[scale=0.20,clip]{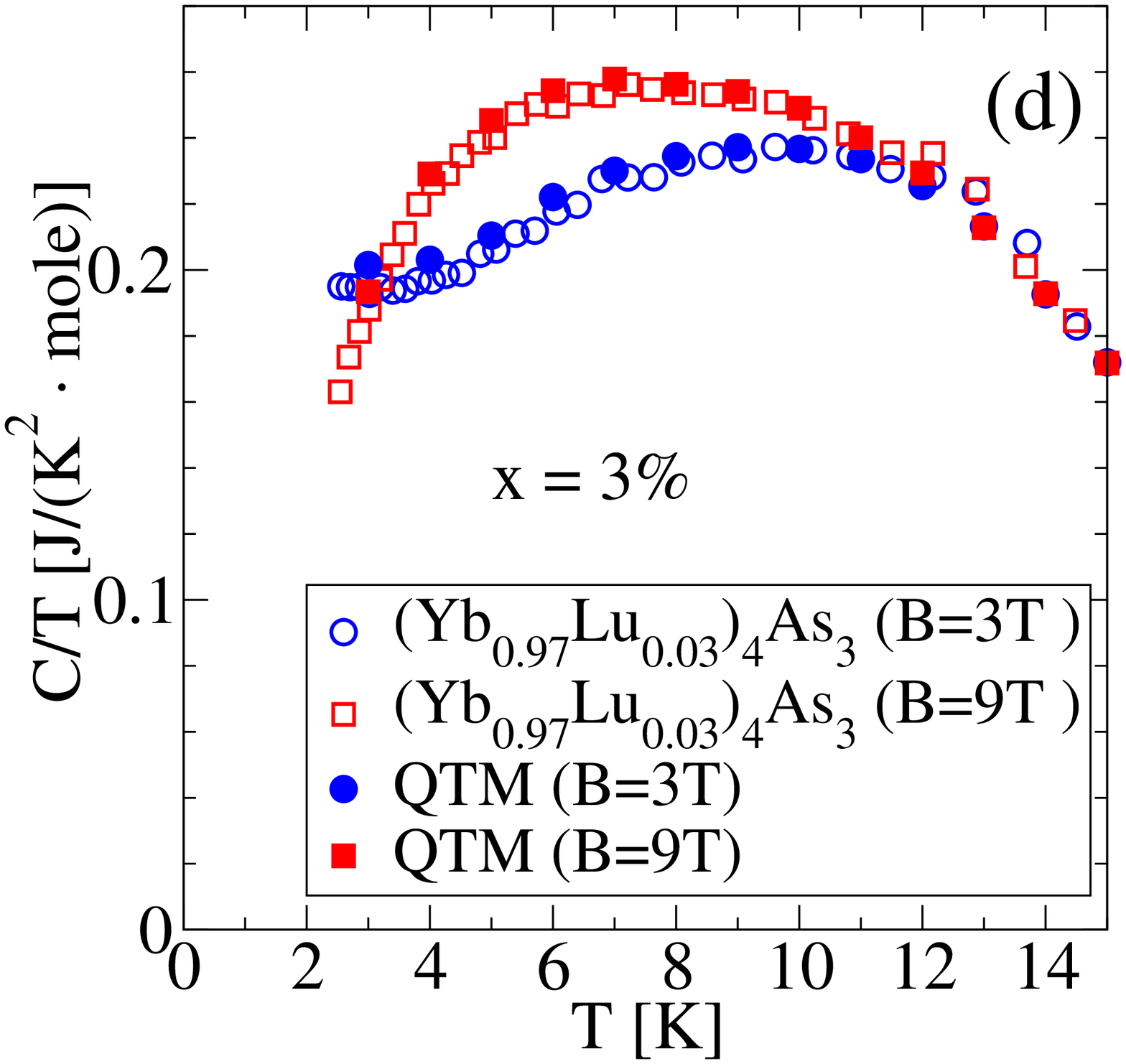}
\end{center}
\caption{(Color online) Comparison between experiment and theory for the diluted
samples subject to an applied field. In panels (a) and (b), the field
applied is
$B$= 12~T and 18~T, respectively. In panels (c) and (d),
the non-magnetic impurity concentration amounts to $x=1$\% and $x=3$\%, respectively.
The symbols are explained in the legend.}
\label{fig2}
\end{figure}

The molar magnetic specific heat data for the diluted (Yb$_{1-x}$Lu$_x$)$_4$As$_3$ 
are plotted in Fig.~\ref{fig2}
and compared with the results of our numerical simulations. The raw outcome of the measurements and the phonon contribution are plotted in Figs.~6-9 in SM.
In Figs.~\ref{fig2}a and \ref{fig2}b the strength of applied field 
$B_{\perp}$ is fixed, whereas in Figs.~\ref{fig2}c and \ref{fig2}d, 
the impurity concentration is kept constant. 
The outright agreement between experiment and theory
provides compelling evidence for the high quality of the model
(\ref{hamilt})
and enhances confidence in our procedure (\ref{specific_heat}) which enables to extract $C_{\perp}$ from $C_{\text{m}}$, having calculated $C_{\parallel}$. In SM, the $C_{\parallel}$ and $C_{\perp}$ contributions to $C_{\text{m}}$ are plotted in Figs.~10-13 and again in the low temperature region they dominate over $C_{\text{ph}}$ so that uncertainties in $C_{\text{ph}}$ have negligible impact on the accuracy of the  $C_{\perp}$ extracted. 

\begin{figure}
\begin{center}
\includegraphics*[scale=0.20,clip]{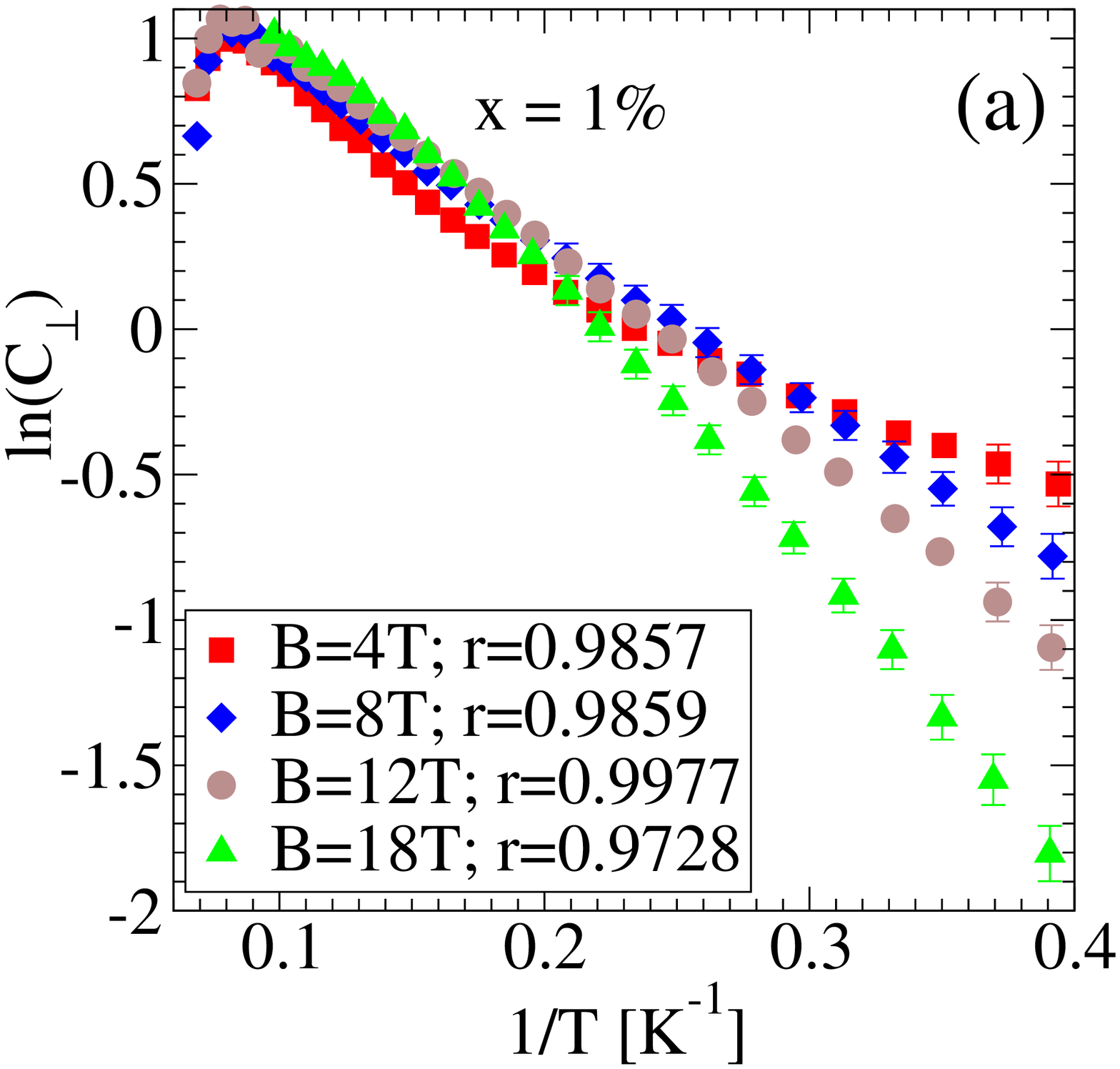}
\includegraphics*[scale=0.20,clip]{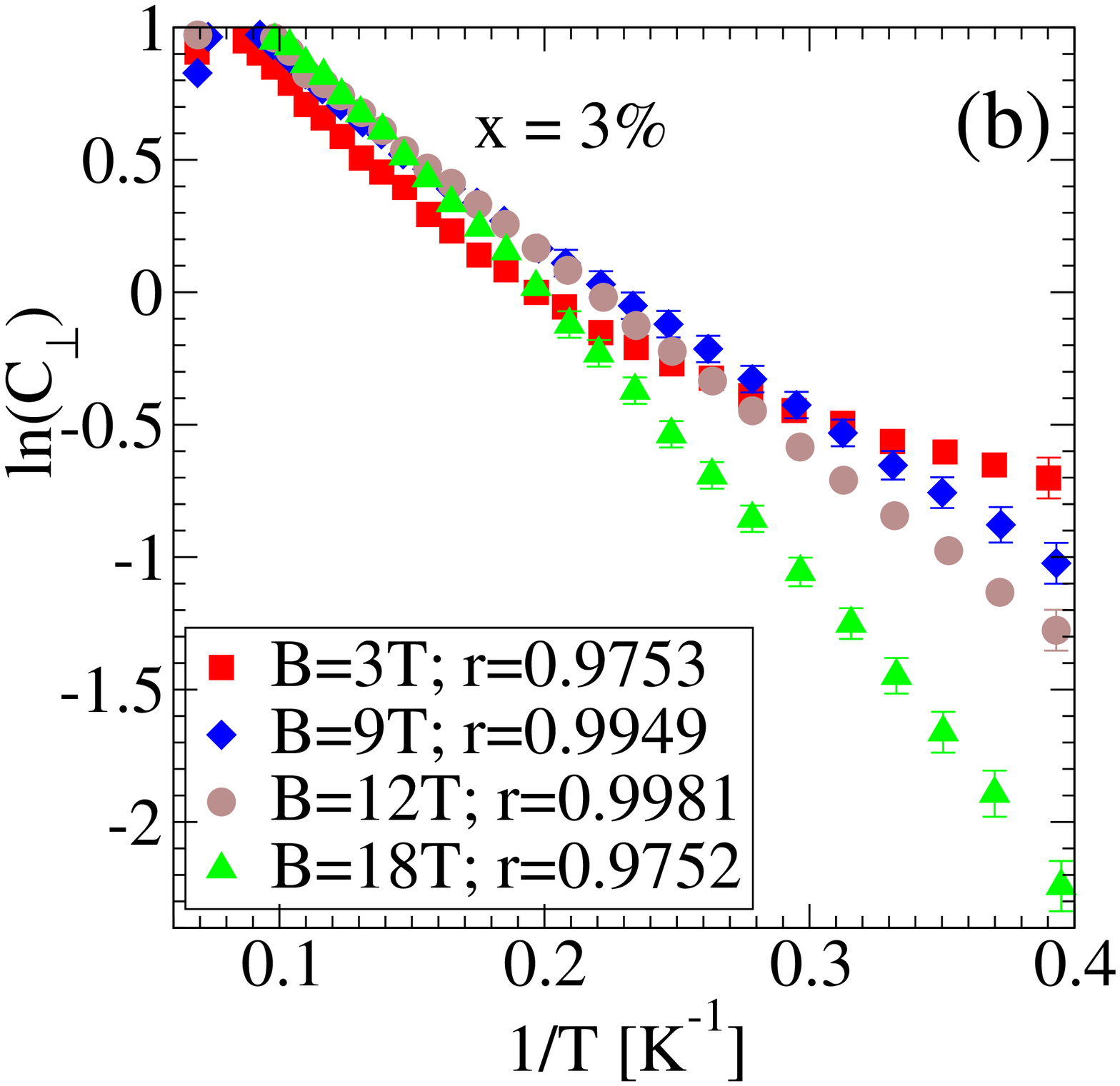}
\includegraphics*[scale=0.20,clip]{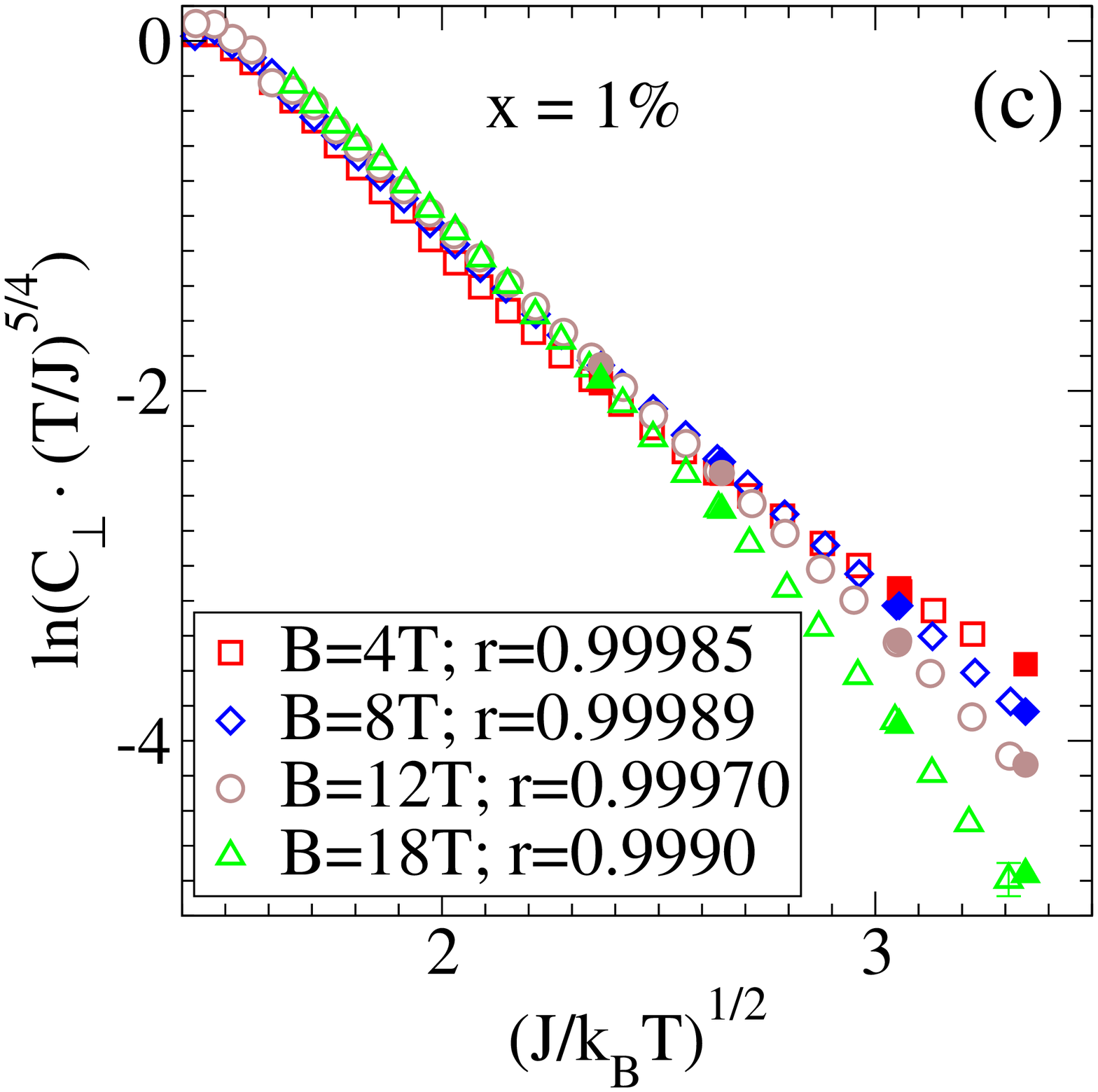}
\includegraphics*[scale=0.20,clip]{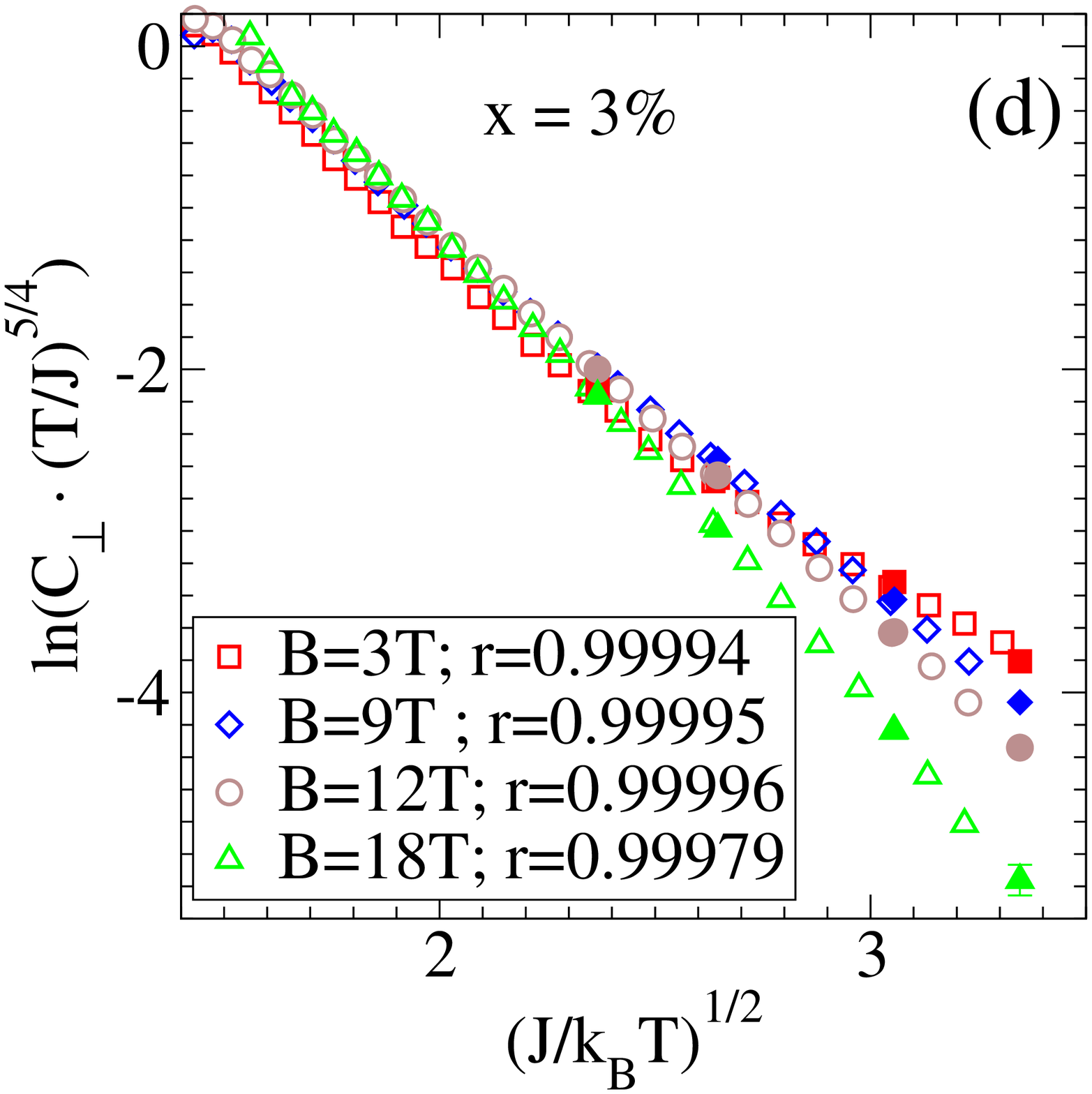}
\end{center}
\caption{(Color online)
The behavior of the contribution $C_{\perp}$
to the specific heat for the diluted (Yb$_{1-x}$Lu$_x$)$_4$As$_3$. 
In the region of low temperatures,
panels (a) and (b) as well as their insets illustrate
non-exponential decay of the $C_{\perp}(T)$ vs. $1/T$.
Panels (c) and (d) 
display the realization of the corresponding BG scaling law of $C_{\perp}(T)\cdot (k_BT/J)^{5/4}$ for $(J/k_B T)^{1/2} \ge 2.4$. 
The open and full symbols represent the experimental data and
the QTM estimates, respectively, and the values {\it r} denote the corresponding
Pearson correlation coefficients.}
\label{fig3}
\end{figure}

To ascertain
that the specific heat behavior characteristic for the BG system
is obeyed in the presence of magnetic field, only the part $C_{\perp}(T)$ of the experimental specific heat shown in Fig.~\ref{fig3} should be considered.
The relevant results
are plotted  in molar units
in Fig.~\ref{fig3} and contain error bars if they exceed the size of the symbols.
In panels (a) and (b)
we demonstrate that
deviations of the specific heat from the exponential decay are stronger than those in Figs.~\ref{fig1}b,
yielding clearly lower values of $r$ ($r \le 0.9981$).
This non-linear dependence agrees with the gapless nature of the Bose glass which is distinguished by a non-exponential decay of the specific heat.

\begin{figure}
\begin{center}
\includegraphics*[scale=0.20,clip]{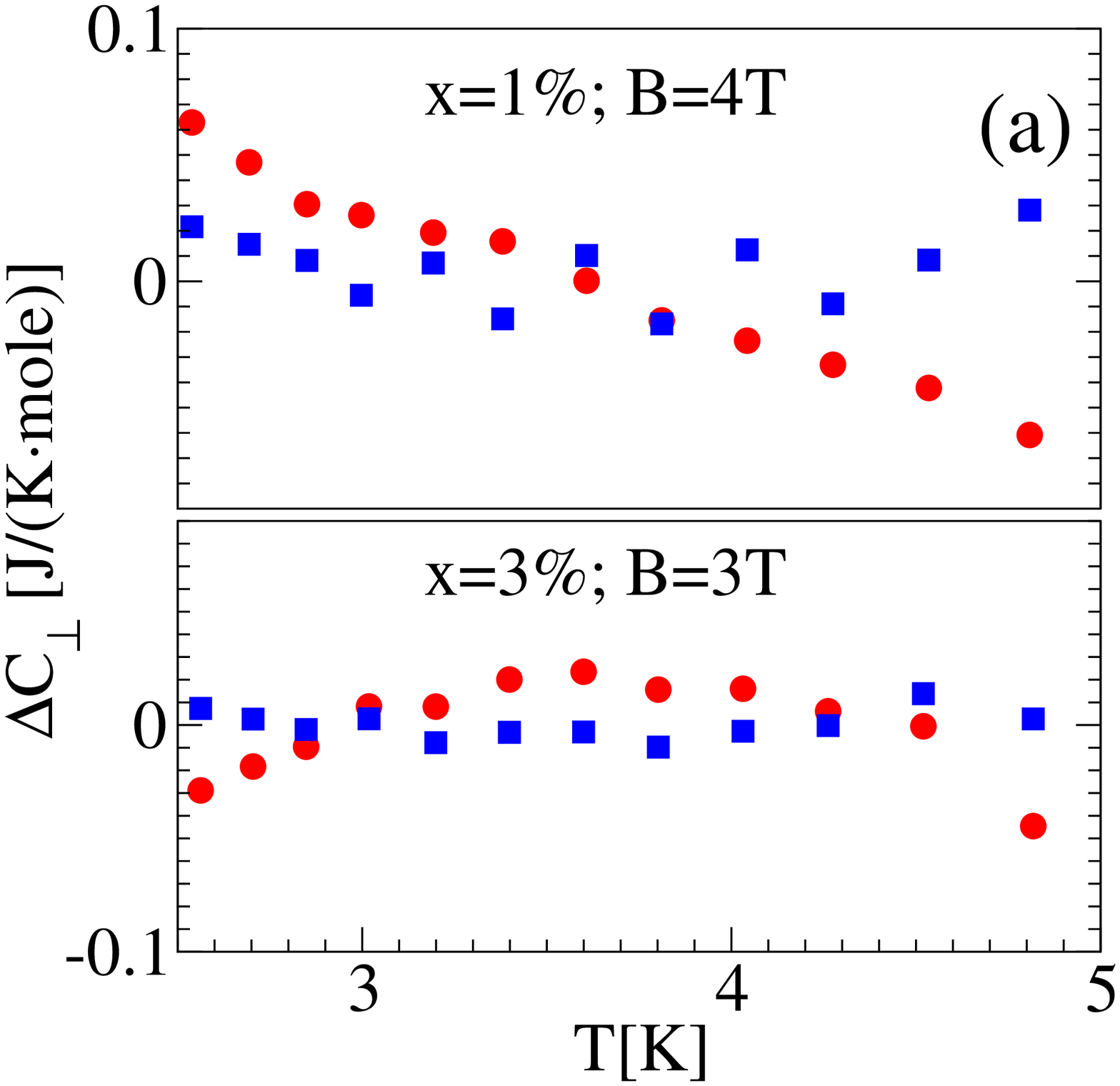}
\includegraphics*[scale=0.20,clip]{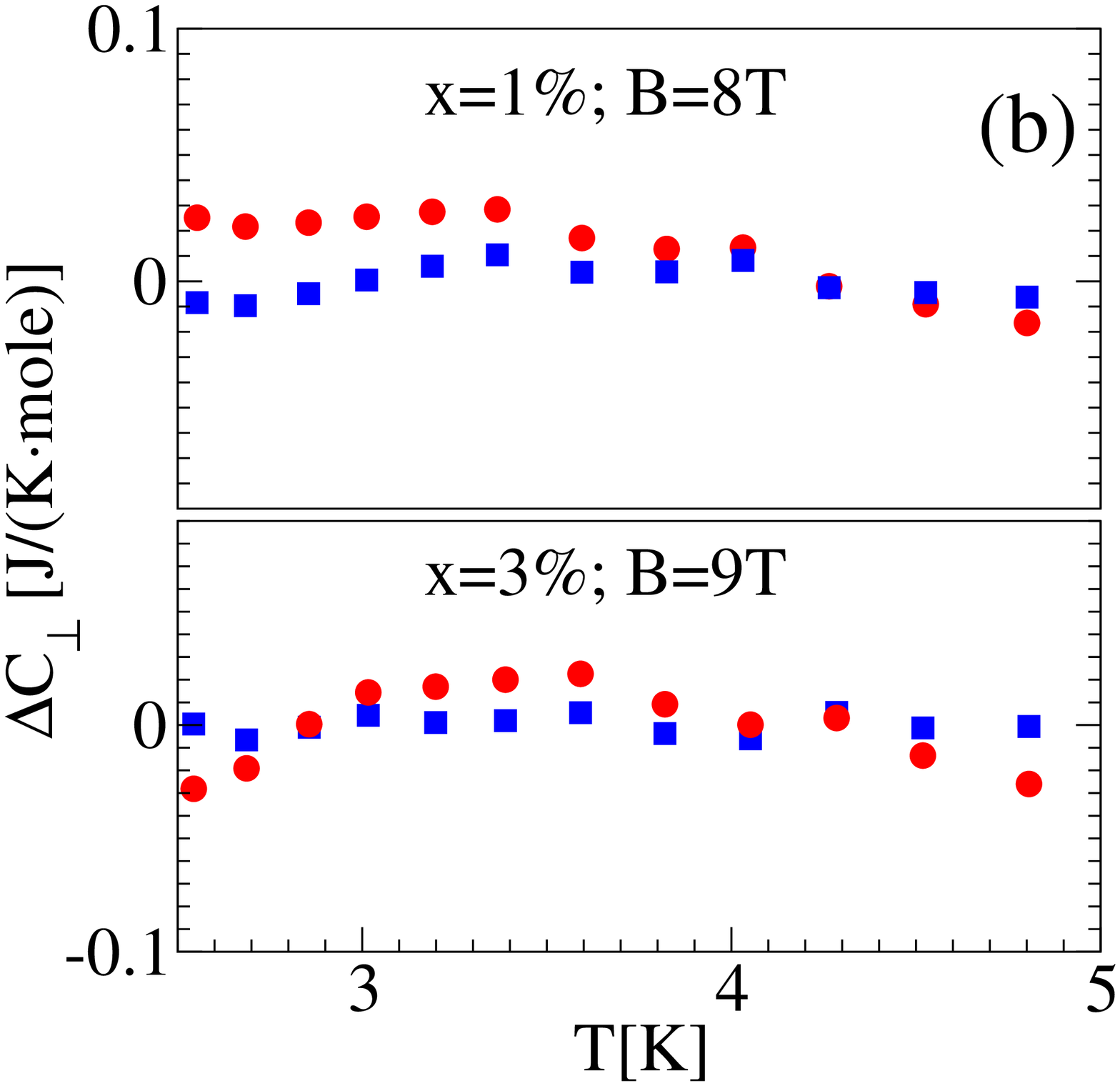}
\end{center}
\caption{(Color online). The deviations of the measured $C_{\perp}$ 
data from those fitted by the
exponential and the stretched exponential dependence plotted 
by full circles and squares, respectively. In panels (a) and (b) some 
representative weak and intermediate applied fields are selected. Error bars attain at most the size of the symbols and are omitted.
}
\label{fig4}
\end{figure}

In panels (c) and (d) of Fig.~\ref{fig3} the rescaled $C_{\perp}$ data
as a function of the variable $(k_B T/J)^{-1/2}$ are plotted and the Pearson coefficients are extracted. The pronounced linear dependence ($0.9990 \le r \le 0.99996$) in the low-temperature region $T \le 5.0$~K (i.e. $(k_B T/J)^{-1/2} \ge 2.4$) points to the stretched exponential behavior.

To check the reliability of our conclusions, we have digitalized the specific heat data plotted in Fig.~2d and Fig.~4 for DTN \cite{YuNature2012}. In the most favorable case $B=0$ we found the coefficients $r=0.99864$ for $x=0$ and the expected exponential decay (in the region $T^{-1}  > 0.96$), and $r=0.99976$ for $x=0.08$ and the expected stretched exponential dependence (in the interval $0.97 < T^{-1/2} < 2.5$, {\it i.e.} neglecting the extreme point lowering $r$). These coefficients characterizing the linear relationship are very close to the corresponding numbers calculated in our study of the Bose glass behavior in (Yb$_{1-x}$Lu$_x$)$_4$As$_3$.

An additional evidence for the BG scaling of the specific heat is provided by the residual values defined as the differences between the experimental data $C_{\perp}$ and the corresponding fits based on 
the exponential (Figs.~\ref{fig3}a,~\ref{fig3}b) and the stretched exponential (Figs.~\ref{fig3}c,~\ref{fig3}d) dependence. The corresponding curves in Fig.~\ref{fig4} are qualitatively different. 
Much smaller values and a random scattering of the residuals obtained for the stretched exponential decay signal the non-exponential behavior. This observation agrees with the similar outcome found for the DTN data (see Figs.~14 and 15 in SM). We note that for pure Yb$_4$As$_3$ the exponential dependence shown in Fig.~\ref{fig1}b is superior to the stretched exponential counterpart as far as the Pearson coefficients and residuals are concerned (see Figs.~5 and 16-17 in SM).

To provide some evidence for the finite uniform susceptibility in (Yb$_{1-x}$Lu$_x$)$_4$As$_3$, we recall both the experimental magnetization data on DTN (Fig.~2a in Yu et al.~\cite{YuNature2012}) as well as the magnetic measurements \cite{Iwasa,Gegenwart2002,H-Aoki} performed on the pure Yb$_4$As$_3$ and the theoretical results for the model (\ref{hamilt}) obtained in some particular cases \cite{Haas,Affleck1999}. For the pure and doped DTN (Fig.~2a in Yu et al.\cite{YuNature2012}) the magnetization profiles nearly coincide and the latter implies a finite uniform magnetic susceptibility characteristic for the Bose glass. Likewise, the dependence of the uniform magnetization of Yb$_4$As$_3$ on the transverse magnetic field (Fig.~4 in Iwasa et al.\cite{Iwasa}) implies the finite susceptibility. We can expect that the susceptibility remains finite under doping, as the local-gap model explaining the exponential suppression of magnetization in DTN \cite{TRoscilde,YuNature2012} is applicable for our system \cite{Haas}, too. We also note that the simplified model (\ref{hamilt}) with $B^x=B^z=0$ is akin to that describing a disordered boson chain in the limit of a large charging energy \cite{Altman}, where the BG phase was also established on the basis of the proper susceptibility behavior.

In conclusion, we have measured the specific heat in the site-diluted (Yb$_{1-x}$Lu$_x$)$_4$As$_3$ which is an ideal model compound
to develop the BG behavior in the transverse magnetic field with respect to the chain direction.
The relevant part $C_{\perp}$ of the
specific heat is found to behave as predicted for a Bose-glass phase, in that it exhibits a non-exponential decay as a function of $1/T$ and obeys the proper BG scaling law. 

P.G. acknowledges discussions with Andreas Ho-
necker. This work was supported by the Polish National Science Centre grant N519 579138 and it was granted access to the HPC 
resources in PSNC Pozna\'n (Poland) and those in Cyprus available within 
DECI program by the PRACE-3IP project No. FP7 RI-312763.


\section*{SUPPLEMENTAL MATERIAL}

The raw field-dependent specific heat measured on the
single-crystal polydomain sample of the site-diluted 
compound (Yb$_{1-x}$Lu$_x$)$_4$As$_3$ and the lattice contribution established earlier are provided to obtain the magnetic part. The latter is split into the inputs arising from the chains parallel and perpendicular to the applied field.
The specific heat residuals calculated for the DTN compound are plotted as a function of temperature to obtain a benchmark for discrimination between the exponential or stretched exponential behavior of the data. Assuming the stretched exponential dependence, the Pearson correlation coefficients are also calculated for the specific heat of the pure Yb$_4$As$_3$ subject to the applied transverse field, to enhance the evidence for the exponential decay.

The field-dependent specific heat measurements 
on the single-crystal (Yb$_{1-x}$Lu$_x$)$_4$As$_3$
with the doping concentration $x=1\%$ or $x=3\%$ were performed, using the same
samples and equipment as those in the previous study
\cite{prb2013rm}. The raw data $C_{\text {exp}}$ for the diluted system and the lattice contribution $C_{\text {ph}}$ obtained earlier~\cite{prb2009rm} for the pure Yb$_4$As$_3$ are plotted in Figs.~5-8 by the symbols and the continuous line, respectively. 
The latter is given~\cite{prb2009rm} by the expression
\begin{equation}
\label{Cph}
C_{\rm ph}=\alpha T^3+\beta T^5
\end{equation} 
with 
$\alpha=1.11\times 10^{-3}$~J/(mol$\cdot$K$^4$) and 
$\beta=4.9\times 10^{-6}$~J/(mol$\cdot$K$^6$). 
Note, that we display the data per mole 
(Yb$_{1-x}$Lu$_x$)$_4$As$_3$, i.e. the heat capacity has not been rescaled 
to the amount of magnetic sites in the system. In addition, we reckon that
the phonon part is unaffected by the field applied and the substitution of the Yb by Lu ions.
In this way
the magnetic specific heat $ C_{\text{m}}$ can be obtained
from the measured specific heat $C_{\text {exp}}$ by extracting 
the lattice contribution $C_{\text {ph}}$ given in Eq.~(\ref{Cph}). For comparison, the temperature dependence of $C_{\text {exp}}$ for the pure sample  is given in Fig.~5.

\begin{figure}
\begin{center}
\includegraphics*[scale=0.25,clip]{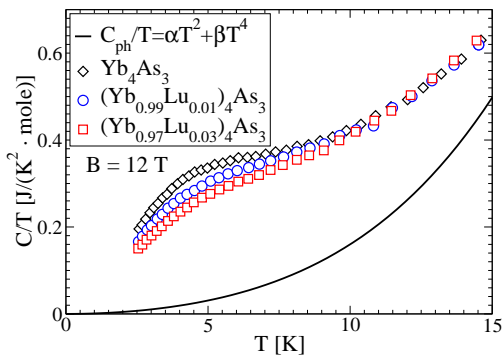}
\end{center}
\caption{(Color online) Raw specific heat as a function of temperature measured on the diluted sample in the applied field $B=12$~T, expressed in the molar units. The phonon part $C_{\text {ph}}$ is plotted by continuous line. The symbols are explained in the legend.}
\label{fig5}
\end{figure}

The raw data $C_{\text {exp}}$ surpass the phonon part $C_{\text {ph}}$ in the entire temperature region spreading up to $T=15$~K and the values $ C_{\text{m}}$ are significantly higher than the uncertainties in the estimates of $C_{\text {ph}}$. Therefore the magnetic part of the specific heat is very accurately established. We remind that the same subtraction procedure was exploited, analyzing the non-magnetic impurity effects in the absence of the applied field~\cite{prb2013rm} and the excellent agreement between theory and experiment was achieved for the magnetic part $C_{\text{m}}$ of the specific heat without any adjustable parameters.

\begin{figure}
\begin{center}
\includegraphics*[scale=0.25,clip]{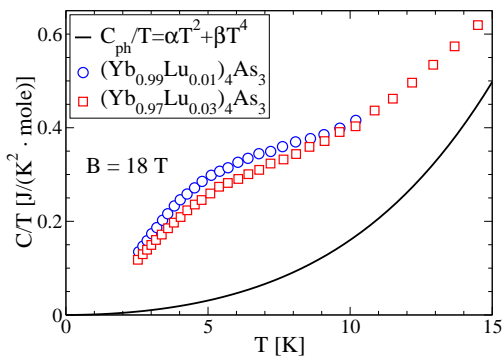}
\end{center}
\caption{(Color online) Raw specific heat as a function of temperature measured on the diluted sample in the applied field $B=18$~T, expressed in the molar units. The phonon part $C_{\text {ph}}$ is plotted by continuous line. The symbols are explained in the legend.}
\label{fig6}
\end{figure}

We note that in the limit $x=0$, the similar relation between  $C_{\text{m}}$ and $C_{\text {ph}}$ was established~\cite{prb2009rm} and after a separation procedure, the extracted contribution from the chains perpendicular to the applied field $C_{\perp}$ was considered in the main part of our publication. On the basis of Fig.~1b plotted therein we argued that $C_{\perp}$ decays exponentially as a function of $1/T$ in the low temperature region. As the Pearson correlation coefficients found are not exactly 1, we check here that the non-exponential dependence deteriorates the values of the corresponding coefficients. The curves plotted in Fig.~9 represent the respective stretched exponentials as a function of $T^{-1/2}$ and imply the Pearson coefficients which are stable but smaller than their counterparts calculated for the exponential dependence ($0.99851 \le r \le 0.99995$).

\begin{figure}
\begin{center}
\includegraphics*[scale=0.25,clip]{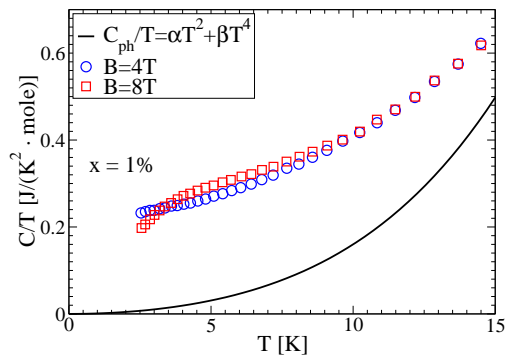}
\end{center}
\caption{(Color online) Raw field-dependent specific heat as a function of temperature measured on the diluted sample for the concentration of the non-magnetic impurities $x=0.01$, expressed in the molar units. The phonon part $C_{\text {ph}}$ is plotted by continuous line. The symbols are explained in the legend.}
\label{fig7}
\end{figure}

The extracted magnetic part of the (Yb$_{1-x}$Lu$_x$)$_4$As$_3$ specific heat is demonstrated as a function of temperature in Figs.~10-12. The curves plotted are the copies of the corresponding counterparts in the main text of the article except for Fig.~10, where the case $x=0$ is included for comparison. However, we also plot here the inputs from the chains parallel and perpendicular to the applied field, {\it i.e.} $\tilde {C_{\parallel}} = 0.25 C_{\parallel}$ and $\tilde {C_{\perp}} = 0.75 C_{\perp}$, respectively. The striking feature of $\tilde {C_{\parallel}}/T$ emerging from Figs.~10-12 is that this part of the specific heat is flat as a function of temperature and very weakly dependent on the field applied or the doping concentration. Compared to $\tilde {C_{\parallel}}/T$, the contribution $\tilde {C_{\perp}}/T$ displays much stronger dependence which is desirable for our analysis.

\begin{figure}
\begin{center}
\includegraphics*[scale=0.25,clip]{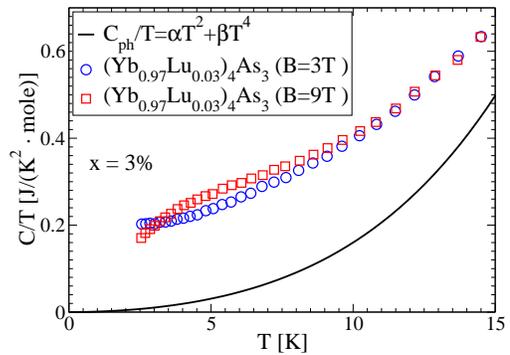}
\end{center}
\caption{(Color online) Raw field-dependent specific heat as a function of temperature measured on the diluted sample for the concentration of the non-magnetic impurities $x=0.03$, expressed in the molar units. The phonon part $C_{\text {ph}}$ is plotted by continuous line. The symbols are explained in the legend.}
\label{fig8}
\end{figure}

In the low temperature region $2.5 \le T \le 5$~K considered, the contribution $\tilde {C_{\perp}}$ dominates over $ C_{\text{ph}}$ and again the uncertainties of $ C_{\text{ph}}$ on the accuracy of $\tilde {C_{\perp}}$ can be neglected. Both the clear separation of $ C_{\text{m}}$ into the inputs $\tilde {C_{\perp}}$ and $\tilde {C_{\parallel}}$, and their prevalence over the $ C_{\text{ph}}$ contribution, provide an evidence for the reliable and accurate determination of the $\tilde {C_{\perp}}$ part of the specific heat.

\begin{figure}
\begin{center}
\includegraphics*[scale=0.25,clip]{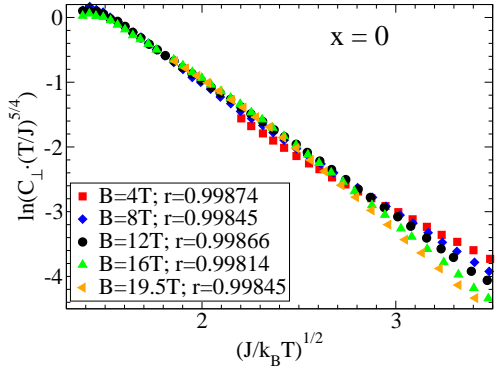}
\end{center}
\caption{(Color online) The field-dependent specific heat $C_{\perp}$ plotted for the pure sample and expressed in the molar units, assuming the stretched exponential dependence. In the legend, the respective correlation coefficients accompany the symbols which describe the curves.}
\label{fig9}
\end{figure}

\begin{figure}
\begin{center}
\includegraphics*[scale=0.25,clip]{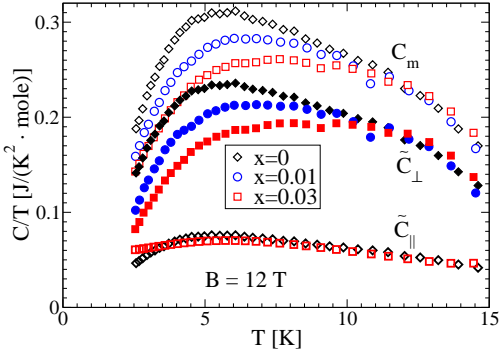}
\end{center}
\caption{(Color online) Magnetic specific heat as a function of temperature measured at $B=12$~T and plotted for the concentrations $x=0$ with diamonds, $x=0.01$ with squares, $x=0.03$ with triangles. The open symbols refer to  $C_{\text{m}}$, whereas the full symbols visualize both $\tilde {C_{\perp}}$ and  $\tilde {C_{\parallel}}$. The flat curves at the bottom demonstrate the $\tilde {C_{\parallel}}$ dependence for the pure sample and that for the doping $x=0.01$ only, to avoid an overlap with the data for $x=0.03$ .}
\label{fig10}
\end{figure}

\begin{figure}
\begin{center}
\includegraphics*[scale=0.25,clip]{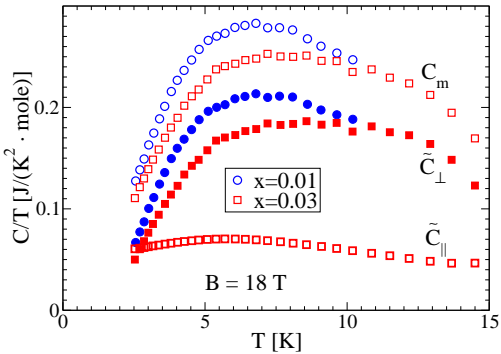}
\end{center}
\caption{(Color online) Magnetic specific heat as a function of temperature measured at $B=18$~T and plotted for the concentrations $x=0.01$ and $x=0.03$. The symbols have the same meaning as those in Fig~\ref{fig5}.}
\label{fig11}
\end{figure}

In our study, the discrimination between the exponential and the stretched exponential behavior of the specific heat $C_{\perp}$ is based on the quantitative criteria, {\it i.e.} the values of the Pearson correlation coefficients and the magnitude of the deviations of the real data from their model dependence obtained from the corresponding fits. The deviations are referred to as the residuals which can be calculated both for the exponential and the stretched exponential relationship. The smaller sizes of the residuals signal the better quality of a fit and help in discrimination between these two types of behavior. 

\begin{figure}
\begin{center}
\includegraphics*[scale=0.25,clip]{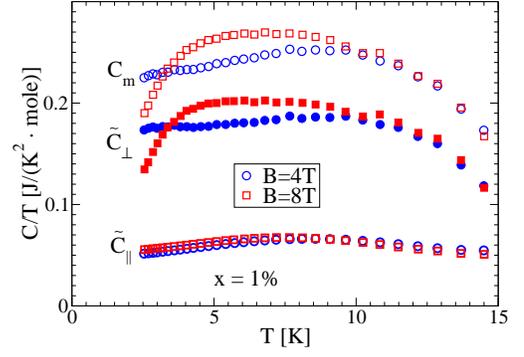}
\end{center}
\caption{(Color online) Magnetic specific heat as a function of temperature measured for $x=0.01$ and a number of the applied fields. The symbols are explained in the legend. For each value of $B$, the quantities $C_{\text{m}}$, $\tilde {C_{\perp}}$ and  $\tilde {C_{\parallel}}$ are plotted.}
\label{fig12}
\end{figure}

For the sake of the quality assessment of our discrimination procedure, the results attained for DTN~\cite{YuNature2012} can be considered as a benchmark. To this end we have digitalized the data presented by Yu et al.~\cite{YuNature2012} in Figs.~2d and~4, we have fitted them by the exponential and stretched exponential dependence and then we have calculated the corresponding residuals. Those obtained in the absence of magnetic field are demonstrated in Figs.~14 and 15. 

\begin{figure}
\begin{center}
\includegraphics*[scale=0.25,clip]{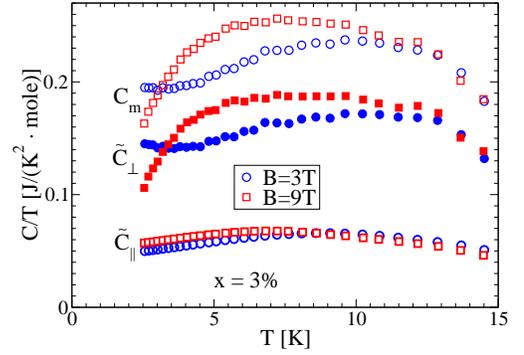}
\end{center}
\caption{(Color online) Magnetic specific heat as a function of temperature measured for $x=0.03$ and a number of the applied fields. The symbols are explained in the legend. For each value of $B$, the quantities $C_{\text{m}}$, $\tilde {C_{\perp}}$ and  $\tilde {C_{\parallel}}$ are plotted.}
\label{fig13}
\end{figure}

For the pure DTN the existing energy gap implies the exponential dependence of the specific heat so that in this case the residuals should be smaller than those calculated for the stretched exponential dependence which is confirmed in Fig.~15. Surprisingly, the value $r = 0.99864$ of the Pearson correlation coefficient for the exponential dependence is only slightly enhanced with respect to $r = 0.99856$ found for the non-exponential analogue so that in both cases the linearity is equally well fulfilled. 

\begin{figure}
\begin{center}
\includegraphics*[scale=0.25,clip]{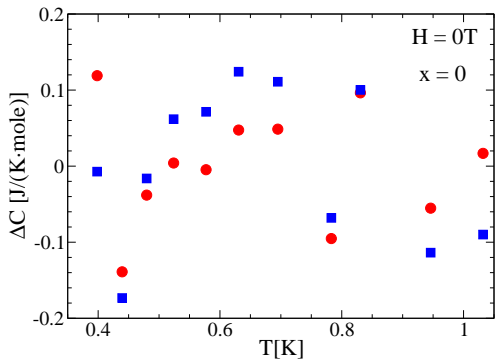}
\end{center}
\caption{(Color online). The residuals of the DTN specific heat with respect to the data arising from the fitting by the
exponential (full circles) and the stretched exponential (squares) dependence for the doping concentration $x=0$.
}
\label{fig14}
\end{figure}

From the other hand, under doping the BG phase sets in and the non-exponential decay supervenes. For that reason the residuals shown in Fig.~15 for the stretched exponential dependence are smaller than their counterparts, whereas the respective correlation coefficients are equal to $r = 0.99976$ and $r = 0.9943$. Summarizing, the appropriate scaling leads to higher values of the Pearson coefficients and to the residuals which are smaller by a factor of 2 or 3 than their counterparts for the alternative dependence. This relation between the residuals and the corresponding values of the correlation coefficients are considered as the benchmarks.

\begin{figure}
\begin{center}
\includegraphics*[scale=0.25,clip]{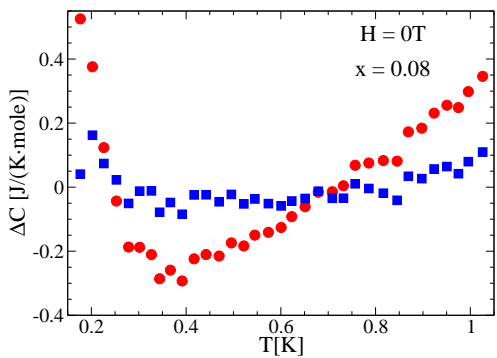}
\end{center}
\caption{(Color online). The residuals of the DTN specific heat with respect to the data arising from the fitting by the
exponential (full circles) and the stretched exponential (squares) dependence for the doping concentration $x=0.08$.
}
\label{fig15}
\end{figure}

We reckon that the highest values of the Pearson correlation coefficients correlated with the smaller residuals provide a criterion for discrimination between the exponential and non-exponential dependence. According to this criterion, the Bose glass behavior is observed in (Yb$_{1-x}$Lu$_x$)$_4$As$_3$. The smaller residuals are found for the curves showing better linearity determined by the corresponding Pearson correlation coefficients (see Fig.~3 in the main part of the work). This feature occurs for the stretched exponential dependence of $C_{\perp}$.

\begin{figure}
\begin{center}
\includegraphics*[scale=0.25,clip]{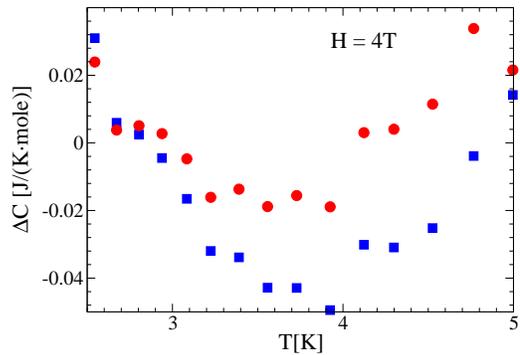}
\end{center}
\caption{(Color online). The residuals of the $C_{\perp}$ specific heat with respect to the data arising from the fitting by the
exponential (full circles) and the stretched exponential (squares) dependence for the pure Yb$_4$As$_3$ and $B=4$~T.
}
\label{fig16}
\end{figure}

Finally we present the residuals calculated for the pure Yb$_4$As$_3$ in Figs.~16 and 17 for the exemplary curves plotted in Fig~1b of the main part of the article and those in Fig.~9. The residuals corresponding to the exponential decay are plotted by full circles and their counterparts by full squares. The former are systematically smaller than the latter and their ratio is consistent with the DTN benchmark. The smaller residuals are correlated with the higher values of the Pearson correlation coefficients and this feature entails the conclusion that it is possible to discriminate the linear exponential decay from its non-exponential counterpart in favor of the former.

\begin{figure}
\begin{center}
\centering
\includegraphics[width=0.8\linewidth,clip]{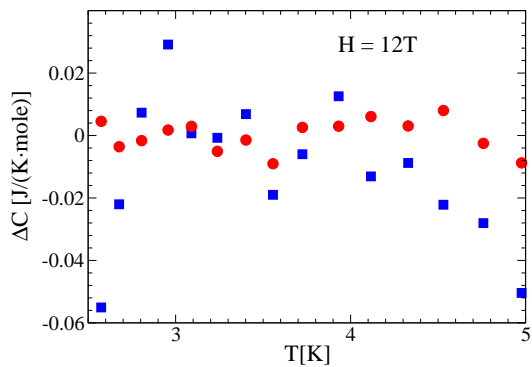}
\end{center}
\caption{(Color online). The residuals of the $C_{\perp}$ specific heat with respect to the data arising from the fitting by the
exponential (full circles) and the stretched exponential (squares) dependence for the pure Yb$_4$As$_3$ and $B=12$~T.
}
\label{fig17}
\end{figure}

\end{document}